# Josephson Junction Field-effect Transistors for Boolean Logic Cryogenic Applications

Feng Wen, Javad Shabani and Emanuel Tutuc, *Senior Member, IEEE*

*Abstract*—Josephson junction field effect transistors (JJ-FET) share design similarities with metal-oxide-semiconductor field effect transistors, except for the source/drain contacts being replaced by superconductors. Similarly, the super current due to proximity effect is tunable by the gate voltage. In this study, we examine the feasibility of JJ-FET-based Boolean logic and memory elements for cryogenic computing, in light of recent advances in novel materials and fabrication techniques. Using a two-dimensional ballistic transport JJ-FET model, we implement circuit level simulations for JJ-FET logic gates, and discuss criteria for realizing signal restoration, as well as fanout. We show that the JJ-FET is a promising candidate for very low power, clocked voltage-level dynamic logic at cryogenic temperatures.

*Index Terms*—Josephson junction, superconducting logic device, ballistic transport[1]

## I. Introduction

REDUCING temperature ($T$) improves several key device metrics of a metal-oxide-semiconductor field effect transistors (MOSFETs). Examples include enhanced channel mobility, and reduced subthreshold swing ($SS$), which scales linearly with $T$ as $SS = 2.3(kT/e)/decade$; $k$ is Boltzmann's constant and $e$ is the electron charge. Channel injection efficiency can be improved in the absence of phonon scattering [1], and the resistance of metal interconnects is expected to decrease at lower temperatures. It is therefore important to investigate under what conditions cryogenic computing provides a net benefit in terms of dissipated energy per switching operation, and whether other devices can offer a benefit in performance at reduced temperatures. For a given technology, with supply voltage $V_{DD}$, the energy dissipated per switching operation is $CV_{DD}^2/2$, where $C$ represents the device capacitance. Assuming $V_{DD}(T)$ can be reduced at lower $T$, by requiring that the total energy dissipated per switching operation, including the cooling cost, does not exceed the room temperature value, we arrive at the following energy balance equation:

$$\frac{CV_{DD}(300\,K)^2}{2} = \frac{CV_{DD}(T)^2}{2} + \frac{300\,K - T}{T} \cdot \frac{CV_{DD}(T)^2}{2} \quad (1)$$

The second term on the right-hand side of Eq. (1) represents the cooling cost at the ideal Carnot efficiency, corresponding to reservoir temperatures of 300 K and $T$. Equation (1) leads to the following simple scaling law for the operating voltage to break even in the ideal cooling limit:

$$V_{DD}(T) = V_{DD}(300\,K)\sqrt{\frac{T}{300\,K}} \quad (2)$$

For example, a $V_{DD}$ value of 0.7 V at room temperature will translate to a break-even value of 83 mV at 4.2 K (He boiling point), which is further reduced to ~26 mV if one factors in realistic cooling efficiencies of 5-10% [2]. As for the gate delay, while delay associated with load and parasitic capacitance being charged/discharged will scale with $V_{DD}$, the transit time delay of channel being switched may not scale down proportionally because of low Fermi velocity at low carrier concentration and reduced thermal excitation of carriers [3]. While the above arguments contain a number of simplifications, they clearly indicate that cryogenic computing using complementary metal–oxide–semiconductor (CMOS) concepts is subject to significant constraints if a net benefit is expected over room-temperature operation with cooling cost factored in. It is therefore highly relevant to critically examine if other devices operating at, or below the break-even $V_{DD}$ value may be used for cryogenic computing applications.

Josephson-logic devices can operate at $V_{DD}$ values in the mV range utilizing superconductivity [4], the most spectacular material property that emerges at low temperatures as a result of electrons forming Cooper pairs. The devices are based on the Josephson junction (JJ), a two-terminal device consisting of two superconductor contacts separated by a weak link, allowing current flow without dissipation (supercurrent) due to proximity effect. The feasibility of Josephson junction field effect transistor (JJ-FET), a three-terminal device with gate-tunable supercurrent, for logic operation was discussed decades ago [5, 6]. The JJ-FET design is similar to a MOSFET except that the source and drain are superconducting at cryogenic temperatures, and the channel length is sufficiently short to allow coherent transport of Cooper pairs through the channel. JJ-FETs have been experimentally demonstrated on various material platforms including Si, Ge and III-V compounds [7-9]. Advances in fabrication techniques and emerging channel materials such as nanowires, III-V quantum-wells, and graphene, as well as the development of transparent semiconductor/superconductor interfaces render the topic of JJ-FETs timely [10-13].

In this study, we address the feasibility of JJ-FET for digital applications. We employ a JJ-FET device model that allows gate-controlled ballistic and coherent transport of Cooper pairs through the channel to examine criteria for signal restoration of several logic gates. We present the results of transient analysis of JJ-FET logic gates, evaluate the impact of fanout on device behavior and discuss various design considerations.

This work was supported by the National Science Foundation Grant No. DMR-1507654, and Intel Corp.

F. Wen, and E. Tutuc are with the Microelectronics Research Center, Department of Electrical and Computer Engineering, the University of Texas at Austin, Austin, TX 78758 USA (e-mail: wenfeng@utexas.edu).

J. Shabani is with Department of Physics, New York University, New York, NY 10003 USA

## II. JJ-FET DEVICE MODEL

We begin by introducing a device model for the JJ-FET. The device, schematically represented in Fig. 1, has the following key length scales: the channel length ($L$), the Cooper pair coherence length ($\xi_0$), and Cooper pair mean free path ($\lambda$). Depending on the interplay between $L$, $\xi_0$ and $\lambda$, the JJ can operate in either short or long ballistic, or diffusive regimes. The Josephson junction is short if $L < \xi_0$, or long if $L > \xi_0$. The Cooper pair transport is ballistic if $L < \lambda$, or diffusive if $L > \lambda$.

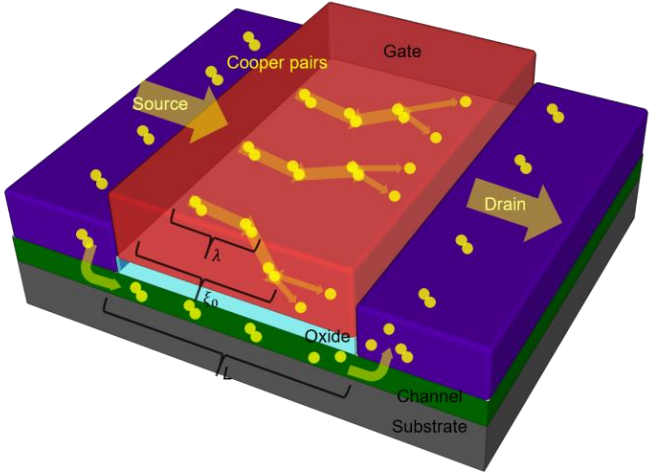

Fig. 1. Schematic of a JJ-FET, consisting of superconducting source and drain, metal gate, oxide, semiconductor channel, and insulating substrate. The length scales $L$, $\xi_0$, and $\lambda$ are indicated. The scattering or destruction of Cooper pairs will occur if $L$ is greater than $\lambda$ or $\xi_0$, respectively, as shown in the figure. The arrows indicate the Cooper pairs or electrons motion.

We consider here the case of a short ballistic JJ device, which satisfies:
$$V_0 = I_C \cdot R_N = \pi \Delta \quad (3)$$
where $I_C$ is the critical current, and $R_N$ the normal resistance of the JJ; and $\Delta$ is the superconductor gap voltage [14]. The conductance of a two-dimensional ballistic transport layer, divided by the average velocity along the channel direction, is
$$g_n = W \cdot \frac{2e^2}{h} \cdot \frac{2}{\pi} \cdot \sqrt{2\pi n} \quad (4)$$
where $W$ is the device width, $h$ is Planck constant, and $n$ the carrier density. The carrier density can be related to gate capacitance ($C_G$), and gate voltage ($V_G$) via $n = C_G(V_G - V_T)/e$; $V_T$ is the threshold voltage. Using Eqs. (3) and (4):
$$I_C = g_n \cdot V_0 = 2V_0 W \cdot \frac{2e}{h} \cdot \sqrt{\frac{2eC_G(V_G - V_T)}{\pi}} \quad (5)$$
It is instructive to introduce an equivalent conductance for the $I_C$ dependence on $V_G$ as:
$$\beta = \frac{dI_C}{dV_G} = V_0 \frac{dg_n}{dV_G} = V_0 W \cdot \frac{2e}{h} \cdot \sqrt{\frac{2eC_G}{\pi(V_G - V_T)}} \quad (6)$$
If we assume the device is in the overdamped limit, where I-V characteristics are non-hysteretic [15, 16], the static I-V characteristics are relatively simple
$$V_{DS} = R_N \sqrt{I_{DS}^2 - I_C^2} \quad (I_{DS} > I_C)$$
$$V_{DS} = 0 \quad (I_{DS} < I_C) \quad (7)$$
where $V_{DS}$ is the voltage drop across the drain and source contacts, and $I_{DS}$ is the drain current. Because Nb and Al are two commonly used superconductors [17, 18], we consider here the cases where the source/drains consist of either Nb or Al, with $\Delta$ values of 1.5 mV or 0.22 mV, respectively [19]. Figure 2 shows the static I-V characteristics of JJ-FETs with the two metal contacts.

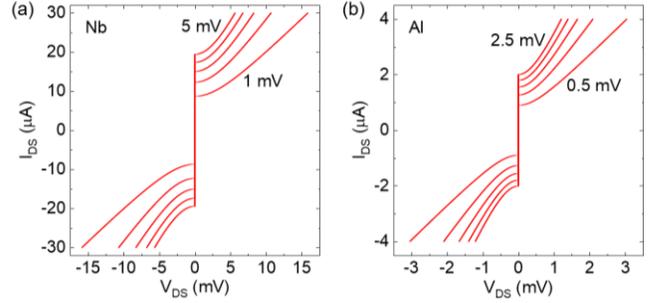

Fig. 2. Calculated I-V characteristics for JJ-FETs with $W = 1$ μm, effective SiO$_2$ oxide thickness of 1 nm ($C_G = 3.45$ μF/cm$^2$) using (a) Nb, and (b) Al contacts. The $V_G - V_T$ values indicated in the figure are changed in 1 mV [panel (a)], and 0.5 mV [panel (b)] increments.

A critical question for a logic gate is if the output voltage is sufficiently large to switch the next stage. To address this question let us assume the output of one JJ-FET is directly driving the gate of a second JJ-FET, whose $I_C$ value, in turn, needs to be sufficiently modulated for a switch. Equation (7) indicates that $V_{DS}$ will be of the order of $V_0$ if $I_{DS}$ is comparable to $I_C$ [4]. The relative change in $I_C$ corresponding to a gate swing of $V_0$ is [20]:
$$\alpha_R = \frac{\beta V_0}{I_C} = \frac{V_0}{2(V_G - V_T)} \quad (8)$$
To achieve signal restoration $\alpha_R \sim 1$, hence $V_G - V_T$ needs to be comparable to $V_0$. This is an intrinsic requirement of JJ-FET logic gate, independent of device scaling and geometry. Figure 3 shows the plots of $\alpha_R$ against $V_G - V_T$ for both Nb- and Al-contact JJ-FETs.

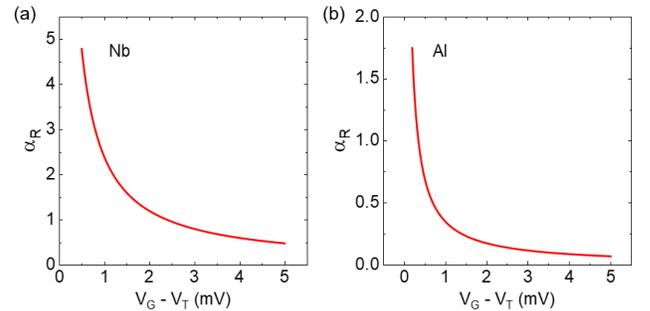

Fig. 3. Calculated $\alpha_R$ vs. $V_G - V_T$ for (a) Nb- and (b) Al-contact JJ-FET.

## III. JJ-FET LOGIC GATES

### A. Static Analysis

In this section we consider logic gates based on JJ-FETs. Figure 4(a-c) illustrate the schematics of JJ-FET inverter, NOR gate and SRAM, respectively. A NOR gate is universal and can be the building block for any combinatorial logic circuit. In this study, we assume all logic devices are biased with ideal DC current ($I_{Bias}$) sources for simplicity. The logic gates operate as follows. For a JJ-FET inverter [Fig. 4(a)], when the input voltage at the gate ($V_{IN}$) is at logic low ($V_{G,Lo}$), ideally zero, the corresponding $I_C < I_{Bias}$, and the JJ-FET is resistive, leading to a finite output voltage ($V_{OUT}$). On the other hand, when $V_{IN}$ is at logic high ($V_{G,Hi}$), the corresponding $I_C > I_{Bias}$, the JJ-FET is superconducting, and $V_{OUT}$ is zero. Signal restoration indicates the finite $V_{OUT}$ at $V_{IN} = 0$ at least equal to $V_{G,Hi}$ to drive the input of the next stage, i.e. same input/output swing. Similarly, in a JJ-FET NOR gate [Fig. 4(b)], the sum of $I_C$ of the two JJ-FETs is smaller than $I_{Bias}$ only when both inputs are at logic low, leading to a finite $V_{OUT}$, and zero otherwise. Connecting two JJ-FET inverters back to back yields an SRAM cell [Fig. 4(c)], where $V_1$ and $V_2$ will a reach stable state of complementary values.

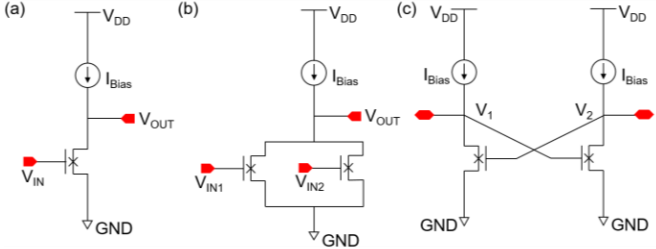

Fig. 4. Schematics of JJ-FET (a) inverter, (b) NOR gate and (c) SRAM.

Next, we investigate the speed of a JJ-FET inverter. In this study, the parasitic capacitance and resistance are ignored. The rising delay ($t_r$) is not expected to be reduced compared to CMOS, because JJ-FET is resistive and the circuit is also an RC network. Moreover, JJ-FET will also draw additional current to the resistive component, namely Josephson current ($I_J$), effectively slowing down the charging speed at the output node. On the other hand, $I_J$ will assist to discharge the output node. The falling delay ($t_f$) is therefore expected to decrease. Though the static model may be oversimplified, it sheds lights on the DC operating point, e.g. $V_T$ and $I_{Bias}$ that will produce the same input/output swing. Figure 5(a) presents the plot of minimum total delay $\tau_{min} = \min(t_r + t_f)$ as a function of $|V_T|$ for an Nb-contact JJ-FET inverter. Each data point is calculated by sweeping $I_{Bias}$ from $I_C$ associated with $V_{IN} = 0$ to $V_{G,Hi}$ and constraining equal input/output swing. The $\tau_{min}$ value reaches a minimum at $V_T \approx -0.4\Delta$. An example of JJ-FET inverter transient analysis is shown in Fig. 5(b). It is clear that the falling edge is more linear than exponential, compared to the rising edge, thanks to $I_J$, which remains $\sim I_C$ and does not linearly decrease with $V_{OUT}$ as the resistive counterpart. The $t_r$ and $t_f$ values are 6.4 and 2 ps for an output capacitance $C_L = 10$ fF, chosen as fan-out of four plus parasitic capacitance. The gate capacitance of a JJ-FET with $W = 1$ μm and $L = 50$ nm is 1.7 fF. We note the actual value of $C_L$ is not significant to demonstrate the logic operation if the static I-V model of JJ-FET is applied, because the JJ-FET logic gates characteristics, except delay will not change with $C_L$. Similarly, a NOR gate can be constructed by simply doubling $I_{Bias}$ since $2I_{Bias} < I_C(V_{G,Hi}) + I_C(V_{G,Lo})$.

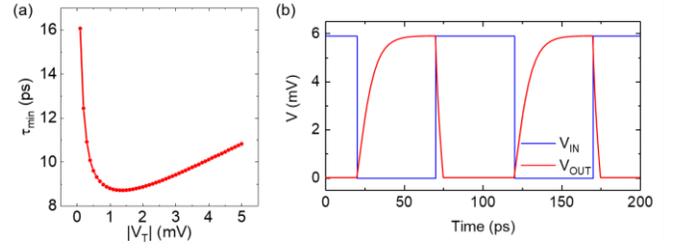

Fig. 5. (a) $\tau_{min}$ vs. $|V_T|$ generated by setting equal input/output swing and sweeping $I_{Bias}$. (b) An example of transient analysis result using the following set of parameters: Swing of input/output = 0 ~ 5.9 mV, $W = 1$ μm, $V_T = -1.2$ mV, $I_{Bias} = 15.1$ μA. The superconducting contacts are Nb.

Applying the same set of parameters of the JJ-FET inverter, Figure 6(a-b) demonstrate the transient response of the JJ-FET SRAM cell with arbitrary initial values of $V_1$ and $V_2$, and the voltage transfer curves, respectively, for Nb-contact device.

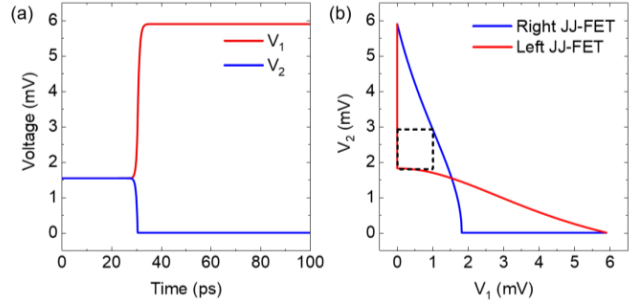

Fig. 6. (a) Transient response of the node voltages at the two outputs. (b) Voltage transfer curves, static noise margin is indicated with a dashed-line-frame box. $C_L = 1$ fF at both outputs. The superconductor contacts are Nb.

### B. Dynamic Analysis

Although we can design the DC operating point of JJ-FET logic gates with the static model, we have made a critical assumption that the JJ-FETs are overdamped in the transient analysis. If a JJ works in the overdamped regime and carries a current exceeding $I_C$, it is non-hysteretic and will become superconducting immediately when the current falls below $I_C$. For an overdamped and resistive JJ-FET, it means the transistor will become superconducting as soon as $V_G$ is increased such that $I_C > I_{DS}$. However, a JJ or JJ-FET is overdamped only when the Stewart-McCumber parameter $Q = 2\pi R^2 C I_C / \Phi_0$ is small ($Q \ll 4$) [15, 16], where $R$ and $C$ are the resistance and capacitance across the junction, and $\Phi_0$ is the magnetic flux quantum. On the other hand, a JJ becomes underdamped if $Q$ is large, which means it remains in the resistive state even when the junction current is reduced below $I_C$, leading to I-V hysteresis. The assumption of overdamped operation becomes questionable if we acknowledge that a logic device has fan-out, and therefore has larger $C$ as well as $Q$ than a single transistor. Moreover, by ignoring the Josephson inductance $L_C = \hbar/2eI_C$, the estimation of gate delay is no longer inaccurate, as the dominant time constant of an overdamped JJ-FET is the Josephson time constant $\tau_{RL} = L_C/R_N = \hbar/\pi e \Delta$. To address these dilemmas, we implement the resistively and capacitively shunted junction (RCSJ) model to describe the device. The I-V characteristics are controlled by the time-dependent Josephson phase $\varphi(t)$, the superconducting contacts phase difference.

$$I_{DS}(t) = C\frac{dV_{DS}(t)}{dt} + \frac{V_{DS}(t)}{R} + I_C \sin\varphi(t)$$
$$V_{DS}(t) = \frac{\hbar}{2e} \cdot \frac{d\varphi(t)}{dt} \quad (9)$$

where $I_C \sin\varphi(t)$ represents $I_J$. We assume the load is purely capacitive and ignore the resistance of the biasing circuitry. Therefore, a JJ-FET inverter's device resistance $R = R_N$ and capacitance $C = C_L + C_J$, where $C_J$ is the junction capacitance assuming gate capacitance splits symmetrically to the source and drain. The gate-tunable $Q$ for a JJ-FET inverter writes:

$$Q_{inv} = \frac{R^2 C I_C}{\Phi_0/2\pi} = \frac{\pi^{3/2} V_0 (C_J + C_L)}{\sqrt{2eC_G(V_{IN}-V_T)}} \quad (10)$$

where $C_J$ and $C_L$ are normalized to $W = 1$ μm of the active JJ-FET. We have $C_L \propto C_G$ if the output is driving inputs of other JJ-FETs, hence for a JJ-FET logic inverter $Q_{inv} \propto \sqrt{C_G/(V_G - V_T)}$.

Figure 7 presents the results of transient analysis for both Nb- and Al-contact JJ-FET inverters, with device parameters adapted from the static analysis based on $V_T \approx -0.4\Delta$. Figure 7(a-b) show the input waveforms for the Nb- and Al-contact JJ-FET inverters, respectively. Figure 7(c-d) show the output waveforms for the Nb- and Al-contact JJ-FET inverters with $C_L = 1$ fF. Similarly, the output waveforms for the Nb- and Al-contact JJ-FET inverters with $C_L = 5$ fF are shown in Fig. 7(e-f). The values of $C_L$ in the dynamic analysis section are chosen to represent the respective operation regime whether hysteresis will latch the JJ-FET in the resistive state or not. Table 1 summarizes the values of $Q$ associated with different $C_L$ and zero $V_{IN}$ for both Nb- and Al-contact JJ-FET inverters. Comparing the output waveforms to those using the static model, superimposed oscillations emerge because of the Josephson effect. Similar to a parallel RLC circuit, the amplitude is attenuated by larger $C_L$, and the frequency is higher for Nb-contact device due to smaller $L_C$. The Nb-contact JJ-FET inverter is significantly faster than its Al-contact counterpart, e.g. $t_f = 0.25$ and $2.5$ ps respectively with $C_L = 1$ fF. The difference in their $t_f$ is comparable to that of their $\tau_{RL}$. If we compare panels (c-d) to (e-f), a distinct difference that can be seen is that JJ-FET inverters fail to be reset into the superconducting states when $V_{IN}$ switches from zero to high when $C_L$ is increased from 1 to 5 fF. Consequently, the finite $Q$ must be accounted for when designing JJ-FET logic gates to avoid an undefined $V_{OUT}$. In the simulation, we find $Q \approx 4$ is the critical value for the Nb-contact JJ-FET inverter to be properly reset, and slightly lower for the Al-contact device as $Q \approx 2.5$. The critical value of $Q$ is closely related to the ratio of $I_{Bias}/I_C(V_{G,Hi})$ due to the hysteretic I-V characteristics. This ratio is 0.5 and 0.7 for the parameter set our Nb- and Al-contact JJ-FET inverter assumes, in agreement with the ratio of return current over $I_C$ corresponding to the critical $Q$ in a hysteretic JJ [16].

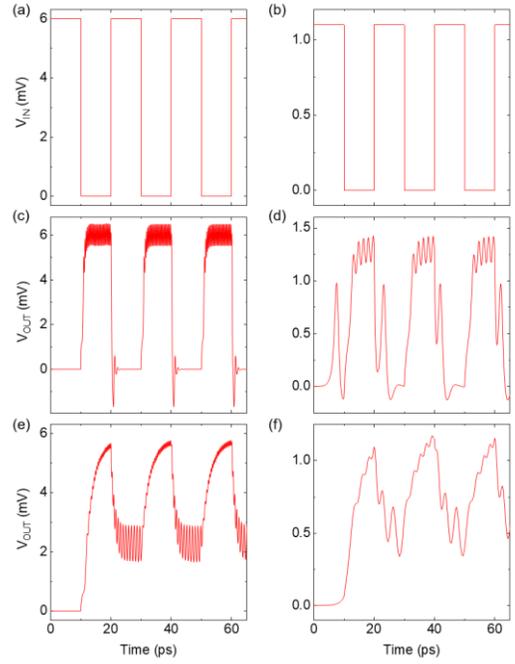

Fig. 7. Square waveforms of $V_{IN}$ for (a) Nb-contact JJ-FET inverter (0 ~ 6 mV) and (b) Al-contact JJ-FET inverter (0 ~ 1.1 mV), period T = 20 ps. Waveforms of $V_{OUT}$ for (c-d) Nb- and Al-contact JJ-FET inverters with $C_L = 1$ fF and (e-f) $C_L = 5$ fF, respectively. For reference, the gate capacitance is 0.7 fF for $W = 1$ μm and $L = 20$ nm. $I_{Bias}$ is slightly reduced to 12 μA from that in the static model for the Nb-contact JJ-FET inverter. $I_{Bias}$ and $V_T$ are 1.06 μA and -0.2 mV for the Al-contact JJ-FET inverter.

TABLE I
$Q$ OF JJ-FET INVERTER

| Contact and $C_L$ | $V_{IN} - V_T = 1.2\ mV$ | $V_{IN} - V_T = 0.2\ mV$ |
|---|---|---|
| Nb, 1 fF | 3.5 | |
| Nb, 5 fF | 15.5 | |
| Al, 1 fF | | 1.0 |
| Al, 5 fF | | 4.4 |

Similarly, we can write $Q$ for a JJ-FET NOR gate as:

$$Q_{NOR} = \frac{R^2 C(I_{C1}+I_{C2})}{\Phi_0/2\pi} = \frac{\pi^{3/2} V_0 (2C_J + C_L)}{\sqrt{2eC_G} \cdot (\sqrt{V_{IN1}-V_T}+\sqrt{V_{IN2}-V_T})} \quad (11)$$

where $I_{C1}$ and $I_{C2}$ are the critical current of the two JJ-FETs. Figure 8 presents the results of transient analysis for JJ-FET NOR gates, with the same device parameters used in JJ-FET inverters and doubled $I_{Bias}$. Figure 8(a-d) show the input waveforms for the Nb- and Al-contact JJ-FET NOR gates, respectively. The signal at $V_{IN2}$ has twice the period and is in phase with that at $V_{IN1}$ to enumerate all four possible logic combinations at the inputs. The output waveforms for the Nb- and Al-contact JJ-FET NOR gates are shown in Fig. 8(e-f) with $C_L = 1$ fF and Fig. 8(g-h) with $C_L = 2.5$ fF. Finally, output waveforms for the Nb- and Al-contact JJ-FET NOR gates with $C_L = 5$ and 7 fF are shown in panel (i) and (j). Table 2 summarizes the values of $Q$ associated with different combinations of $V_{IN1}$, $V_{IN2}$ and $C_L$ for both Nb- and Al-contact JJ-FET NOR gates. Again, the values of $C_L$ and therefore $Q$ are crucial to determine the behavior of the JJ-FET NOR gates. In Fig. 8(e-f), $V_{OUT}$ correctly reproduces the response of a NOR gate, e.g. $V_{OUT}$ only becomes finite when both inputs are low. However, when $C_L$ is increased from 1 to 2.5 fF, $V_{OUT}$ fails to be reset to zero when only one of the inputs switches from low to high and outputs an undefined intermediate state [Fig. 8(g-h)]. When $C_L$ is further increased to 5 and 7 fF for the Nb- and Al-contact JJ-FET NOR gates, $V_{OUT}$ remains finite even when both inputs switch to high. Consequently, there are two undefined intermediate states, as shown in Fig. 8(i-j).

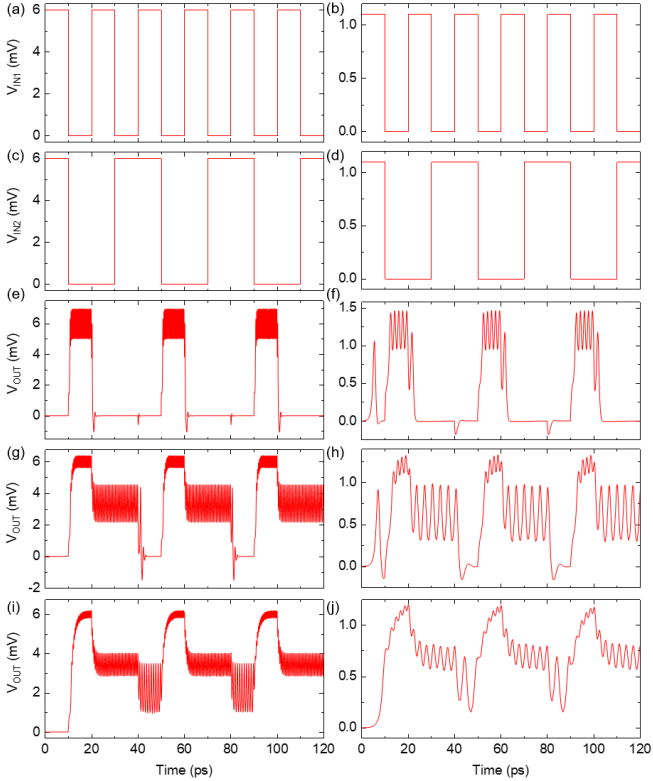

Fig. 8. Square waveforms of $V_{IN1}$ for the (a) Nb-contact JJ-FET NOR gate (0 - 6 mV) and (b) Al-contact JJ-FET NOR gate (0 - 1.1 mV), period T = 20 ps. Square waveforms of $V_{IN2}$ for the (c) Nb- and (d) Al-contact JJ-FET NOR gate, period T = 40 ps. Waveforms of $V_{OUT}$ for the Nb- and Al-contact JJ-FET NOR gate with (e-f) $C_L$ = 1 fF, (g-h) $C_L$ = 2.5 fF, (i) $C_L$ = 5 fF and (j) $C_L$ = 7 fF, respectively. $I_{Bias}$ = 24 and 2 μA for Nb- and Al-contact JJ-FET NOR gates.

TABLE II
$Q$ OF JJ-FET NOR GATE

| Logic values of $V_{IN1}$ and $V_{IN2}$ | 00 | 01 or 10 | 11 |
| --- | --- | --- | --- |
| Nb-contact, $C_L$ = 1 fF | 7.1 | 2.9 | 2.0 |
| Nb-contact, $C_L$ = 2.5 fF | 12.6 | 6.7 | 4.3 |
| Nb-contact, $C_L$ = 5 fF | 21.8 | 12.0 | 8.0 |
| Al-contact, $C_L$ = 1 fF | 2.54 | 1.0 | 0.7 |
| Al-contact, $C_L$ = 2.5 fF | 4.5 | 2.3 | 1.5 |
| Al-contact, $C_L$ = 7 fF | 10.4 | 5.6 | 3.8 |

*C. Global Clock to Reset JJ-FETs*

In the previous section we showed that the $Q$ = 0 approximation is not applicable if the fanout load capacitance is taken into account. The JJ-FETs usually enter the underdamped regime and become hysteretic, at least when inputs are low. On the other hand, JJ-FETs can turn back to overdamped when inputs switch from low to high, thanks to a low $Q$ from optimized design. In this case, we can harness the unique feature of the superconducting logic device, where $t_f$ is determined by $\tau_{RL}$. Thus, a promising design of JJ-FET logic gate resembles the dynamic logic gate, featuring a pre-charge of $V_{OUT}$ and monotonically rising $V_{IN}$ during evaluation.

Voltage-state JJ logic device is not a new concept. Decades ago the transformation of a JJ into switching element was realized through magnetic coupling or current injection into the channel [4]. Those JJs were underdamped with high $Q$ due to the obsolete fabrication technology and capacitor device geometry, and required an ac power supply and other circuit elements to isolate the output from the input. An ac power supply was used since in an underdamped JJ, resetting from the resistive to superconducting state requires the ratio of $I_J/I_C$ to be lowered to a small value for a period of time and this can only be achieved by lowering $I_{Bias}$. On the other hand, the JJ-FET is non-reciprocal and has a low $Q$ thanks to modern transistor fabrication technology and co-planar geometry. Moreover, the JJ-FET has a gate-tunable $I_C$ and can conceivably be reset to superconducting state by increasing $V_G - V_T$. We find that a short voltage pulse can be applied to $V_{IN}$ to temporarily boost $V_G - V_T$ and reset $V_{OUT}$ back to zero. Figure 9(a-b) present the output waveforms for the Nb- and Al-contact JJ-FET inverters with $C_L$ = 5 fF in response to a square waveform a period of 50 ps. Due to the underdamped operation, $V_{OUT}$ remains finite when $V_{IN}$ switches to high. Figure 9(c-d) show the results of $V_{OUT}$ with an addition of clock signals ($V_{CLK}$) with the same period in red and blue lines, respectively. $V_{CLK}$ includes a 5 ps voltage pulse of 10 and 4 mV for the Nb- and Al-contact JJ-FET inverters, which is applied at the rising edge of $V_{IN}$. The pulse width ($t_{PW}$) can be narrower with a larger amplitude for higher speed, e.g. a pulse of 20 mV, 2 ps can reset this Nb-contact JJ-FET inverter. Conversely, $t_{PW}$ of such a transient pulse can be relaxed with a lower operating frequency. A simplification is made that $V_{IN}$ and $V_{CLK}$ add to each other. $V_{OUT}$ is successfully restored to zero with $V_{CLK}$ added. Similarly, Fig. 9(e-f) show the results of the voltage pulse activated at the falling edge of $V_{IN}$, where red and blue lines represent $V_{OUT}$ and $V_{CLK}$. Though the rise of $V_{OUT}$ is delayed, it does reach the target value. Therefore, a global clock mechanism can be introduced if underdamped operation cannot be avoided, at the expense of lower speed. Indeed, since the logic value of $V_{OUT}$ must be evaluated when $V_{CLK}$ = 0, the waveform half-period should exceed $t_{PW} + t_r + t_H$, where $t_H$ is the hold time. It is noteworthy that $V_{CLK}$ may be synchronized with the clock in a dynamic logic gate.

## IV. DISCUSSION

In our analysis throughout we have assumed the JJ-FET gate overdrive $V_G - V_T$ controls the gate charge ($q_G$) and therefore $I_C$ by modulating the weak link in the channel. However, this unidirectional JJ-FET model where $I_{DS}$ is set by $q_G$ without a back reaction is thermodynamically unsound [21]. The energy of a JJ-FET is:

$$E_{JF} = \frac{q_G^2}{2WLC_G} + \frac{\hbar}{2e}I_C(q_G)(1 - cos\varphi(t)) \quad (12)$$

where $I_C(q_G)$ is $I_C$ with gate charge $q_G$. Since $V_G - V_T = \partial E_{JF}/\partial q_G$, we find $q_G$ is reduced, compared to the unidirectional model

$$q_G = WLC_G(V_G - V_T) - \frac{\hbar\beta(q_G)}{2e}(1 - cos\varphi(t)) \quad (13)$$

where $\beta(q_G)$ is $\beta$ with gate charge $q_G$. The back reaction is negligible when the right hand side second term is small, namely $\hbar\beta(q_G)/2eWLC_G(V_G - V_T) \ll 1$, a condition which does not hold for the logic device operation regime discussed here. For the parameters used in the Nb-contact JJ-FET inverter at zero $V_{IN}$, i.e. $V_G - V_T = 1.2$ mV, $W = 1$ μm and $L = 50$ nm, the two terms of Eq. (13) RHS are comparable.

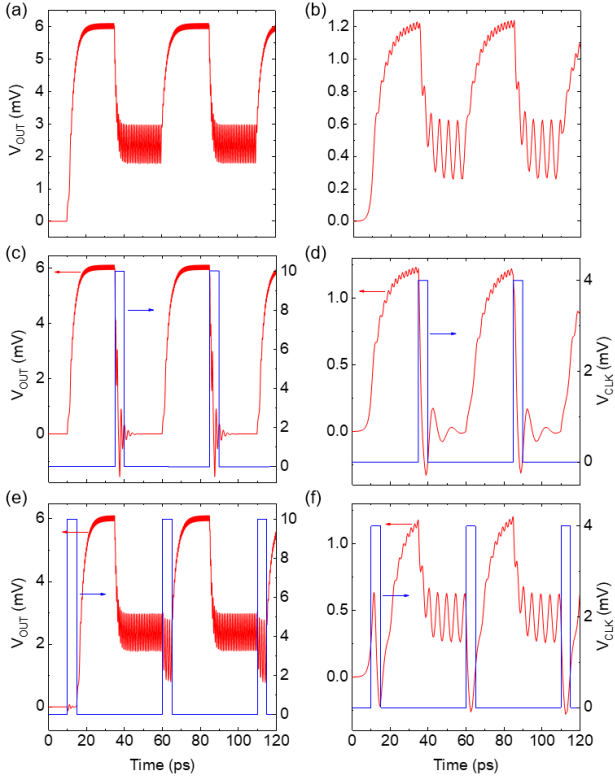

Fig. 9. (a-b) Output waveforms of Nb- and Al-contact JJ-FET inverters without $V_{CLK}$. (c-d) Output waveforms with $V_{CLK}$ switching from low to high at the $V_{IN}$ rising edge. (e-f) Output waveforms with $V_{CLK}$ switching from low to high at the $V_{IN}$ falling edge. $V_{CLK}$ is indicated in blue curves in panels (c-f). The other parameters are the same as those used in Fig. 7.

We note that $\beta$ decreases with increasing $V_G - V_T$; hence, the back reaction becomes particularly important when $V_{IN}$ is low. This implies that if we ignore the back reaction, the channel can be fully depleted and $V_{OUT}$ will be in an undefined state. The back reaction can be compensated by shifting $V_T$ to a more negative value. For example, the operation of an Nb-contact JJ-FET inverter is restored by setting $V_T = -3.5$ mV. While the back reaction adds extra complexity to the design of JJ-FET logic gates, if it is properly compensated at zero $V_{IN}$, at high $V_{IN}$ its impact is reduced.

It is of interest to investigate how advances in nano-fabrication will influence the JJ-FET logic gates. Transistor scaling effectively produces smaller $L$ and larger $C_G$. In order to validate the model used in this study, $L$ needs to be sufficiently short compared to the superconducting coherence length

$$\xi_0 = \frac{\hbar v_F}{\pi e \Delta} = \frac{\hbar^2 \sqrt{2\pi C_G (V_G - V_T)}}{m^* \pi e^{3/2} \Delta} \tag{14}$$

where $v_F$ and $m^*$ are the Fermi velocity and effective electron mass in the semiconductor channel, which are crucial to achieve a significant $\xi_0$ at the low $V_G - V_T$ required by JJ-FET logic gates. A large $C_G$ in the scaled JJ-FET device promises a long $\xi_0$ and small $\gamma$. Alternatively, it may also allow a lower $V_G - V_T$ to reduce power consumption. However, $Q$ will increase consequently, which imposes a design trade-off of $C_G$ and potentially justifies the necessity of $V_{CLK}$.

The choice of superconductor contacts is another important factor. Choosing a low $\Delta$ contact has various benefits. In light of the requirements of finite $V_G - V_T$ for reasonable $\gamma$, $\xi_0$ and $Q$, and appreciable $\alpha_R$ for signal restoration, we usually have $V_G - V_T \sim \Delta$. Then we can approximate the power consumption $P \propto I_C V_0 \propto \Delta^{5/2}$ and power-delay product $P\tau_{RL} \propto \Delta^{3/2}$, indicating that smaller $\Delta$ yields more efficient JJ-FET logic gates. Additionally, we have $Q \propto \sqrt{\Delta}$ and $\xi_0 \propto 1/\sqrt{\Delta}$, promoting the overdamped operation and relaxing the requirement of reduced $L$ and $m^*$. However, we also have $\gamma \propto 1/\sqrt{\Delta}$, which indicates a low $\Delta$ JJ-FET is less immune to the back reaction. Also, a larger $\Delta$ is favored for faster speed given $\tau_{RL} \propto 1/\Delta$. The above arguments impose a design trade-off for $\Delta$.

Although JJ-FET logic gates cannot relax the requirement of ultra-precise control of $V_T$ compared to cryogenic CMOS, they provide a better circuit tolerance in the sense that JJ-FETs are always in the on-state while a CMOS device has to make a transition between on- and off-state within the operating voltage window. Moreover, the speed of the JJ-FET logic gate is limited by $\tau_{RL}$ if designed properly for overdamped operation, as opposed to the $RC$ time constant in a CMOS device, which can be unacceptable high for the low carrier concentrations due to small $V_{DD}$ at cryogenic temperature. On the other hand, JJ-FET logic gates possess a smaller break-even operating voltage than cryogenic CMOS if we factor in the static power consumption. JJ-FET logic gates also demonstrate great compatibility with emerging material platforms, e.g. the III-V quantum-well JJ-FET is a depletion-mode n-type device [11]; the graphene channel can reach its charge neutrality point at a slightly negative $V_G$ [10] and the proposed JJ-FET logic gates can circumvent the issue of low on-off ratio for CMOS [22, 23]. Moreover, the low $m^*$ and high $v_F$ in III-V quantum-well, and especially in graphene mitigate the conflict between long $\xi_0$ and low $V_G - V_T$. e.g. Dirac electrons in graphene have an $\xi_0 = 70$ nm and 470 nm in Nb- and Al-contact JJ-FETs [24], respectively.

V. CONCLUSION

When cooling costs are factored in, the operating voltage of CMOS-based circuits has to be scaled down significantly for cryogenic computing to provide a net power reduction. JJ-FET Boolean logic can harness the superconducting property of these devices at these cryogenic temperatures, and provide low operating voltage on the order of superconductor gap voltage. Assuming a short ballistic transport length, we employ the static and RCSJ model to capture the behavior of JJ-FET logic gates with fan-out. A global clock can mitigate the underdamped operation, if necessary. Transistor scaling and the choice of different superconducting contacts have notable impacts on the device operation. For example, reduced gate dielectric thickness guarantees better back-reaction immunity, but favors underdamped operation, and larger gap voltage ensures a faster operation speed but at the cost of reduced coherence length, hence channel length. We find JJ-FET logic gates can be a promising candidate for dynamic logic elements with ultra-short fall times, and can utilize the advantages of emerging channel materials like III-V quantum wells, and graphene.

ACKNOWLEDGMENT

We thank S. K. Banerjee and R. Pillarisetty for discussions.